\documentclass[twocolumn,nofootinbib]{revtex4-1}

\usepackage{hyperref}  
\usepackage{amsmath}  
\usepackage{amsfonts} 
\usepackage{graphicx} 

\begin{document}


\title{Full non--LTE spectral line formation I. Setting the stage}

\author{Fr\'ed\'eric Paletou} \email{frederic.paletou@univ-tlse3.fr}
\affiliation{Universit\'e de Toulouse, Observatoire
  Midi--Pyr\'en\'ees, Cnrs, Cnes, Irap, F--31400 Toulouse, France}

\author{Christophe Peymirat} \email{christophe.peymirat@univ-tlse3.fr}
\affiliation{Universit\'e de Toulouse, Facult\'e des Sciences et
  d'Ing\'enierie, F--31062 Toulouse cedex 9, France}


\date{\today}

   \begin{abstract}Radiative transfer out of local thermodynamic equilibrium
     (LTE) has been increasingly adressed, mostly numerically, for
     about six decades now. However the standard non--LTE problem most
     often refers to the \emph{only} deviation of the distribution of
     photons from their equilibrium i.e., Planckian,
     distribution. Hereafter we revisit after Oxenius (1986) the
     so-called ``full non--LTE'' problem, which considers to couple and
     therefore to solve self--consistently for deviations from
     equilibrium distributions of photons as well as for massive
     particles constituting the atmospheric plasma.
     \end{abstract}



   \maketitle

%

\section{Introduction}

The ``standard'' non--LTE (NLTE) radiation transfer problem considers that
the distribution of \emph{photons} in a given ``atmosphere'' that is,
more generally, whatever sample of celestial body material where
light--matter interactions are taking place, may depart from the
equilibrium distribution described by the Planck law (Planck 1900).

It is a routinely solved problem in astrophysics which, after about
six decades of constant endeavour, nowadays considers also both
complex atomic models and atmospheric structures (see e.g., the
monograph of Hubeny \& Mihalas 2014). However, the vast majority of
these problems still rely on what became a somewhat \emph{implicit}
assumption that the distribution of \emph{atoms} both responsible for,
and experiencing light--matter interactions remain a priori known and
characterized by the Maxwell--Boltzmann \emph{equilibrium}
distribution. Another issue related with the standard NLTE radiation
transfer problem concerns the \emph{redistribution}, at least in
frequencies if we restrict ourselves to isotropic scattering, of
photons after scattering onto these very massive particles. It is also
important to realize that, in the vast majority of the cases, and even
though the problem of partial frequency redistribution (hereafter PRD)
has been addressed quite early in the 60's (e.g., Avrett \& Hummer
1965, Auer 1968 for pioneering works), the standard NLTE problem also
remains, in most cases, limited to the frame of \emph{complete}
redistribution (hereafter, CRD).

The solution of a \emph{full(er)} non--LTE radiation transfer problem
would consist in assuming that the velocity distribution functions
(hereafter VDF) of massive particles may \emph{also} depart from an
equilibrium distribution, as it is now commonly assumed for the
photons. And therefore the solution of this problem would consist in
the \emph{self--consistent} resolution of a set of kinetic equations
both for massive particles, potentially including free electrons, and
the photons.

This general issue has already been discussed, and summarized mainly
by Oxenius (1986; see also Oxenius \& Simonneau 1994). However, so far
and because of the inherent difficulty of the problem, only a few
studies have already been conducted in the past (see e.g.,
Borsenberger et al. 1986, 1987).

We believe that the last decades of evolution of numerical methods
(see e.g., Lambert et al. 2016 and references therein, as well as
Noebauer \& Sim 2019 for a recent review on Monte--Carlo radiative
transfer) for radiation transfer should allow to reconsider this
issue, and evaluate up to which level of difficulty one could
reasonably be able to achieve nowadays. This new evaluation should
also give us new hints about which kind of astrophysical problems
should be revisited, using full NLTE.

In this study, we set again the various equations governing this
problem, using as much as possible standard notations. We adopt also
several physical assumptions, already discussed in past studies, in
order to be able to set up initial numerical experiments. We discuss
partial vs. complete frequency redistribution, as seen from this
description of the problem. We present indeed some effects expected on
departures of the VDF of excited atoms vs. the Maxwell--Boltzmann
distribution, as well as significative differences between computed
and \emph{radiation dependent} emission profiles, and \emph{a priori
  given} absorption profile, for a simple two-level atom model. We
finally discuss future work in that somewhat forgotten physical frame
for the (unpolarized) radiation transfer problem.

\section{A simplified Boltzmann equation for massive particles}

Let us write the evolution equation for a general
distribution $F(t, \vec{r}, \vec{v})$ as:

\begin{equation}
  L[F] = C[F] \, ,
\end{equation}
where $L$ is the \emph{Liouville} operator, and $C$ the collision
operator.

Therefore, the left-hand side of Eq.\,(1) has the general expression:

  \begin{equation}
    L[F]={{\partial F} \over {\partial t}} + {p \over m}
    {{\partial F} \over {\partial r}} +
    K {{\partial F} \over {\partial p}} \, ,
  \end{equation}
  where after the \emph{time} evolution term, we find an
  \emph{advection} one, and finally a \emph{force} term (where we used
  $K$ to avoid any confusion with the distribution $F$).
  
The right-hand side of Eq.\,(1) describes the various origins of
collisions, which may induce modifications of the distribution. It
is usually expressed as the sum of three contributions such as:

  \begin{equation}
    C[F]=\left( {{\delta F} \over {\delta t}} \right)_{\rm rad.}
    + \left( {{\delta F} \over {\delta t}} \right)_{\rm inel.}
    + \left( {{\delta F} \over {\delta t}} \right)_{\rm el.} \, .
  \end{equation}

As a matter of fact, two main mechanisms are able to produce changes
in a distribution $F$. One of them is the so-called \emph{streaming}
of the massive particles, which move on their trajectories, and a
second one is due to \emph{collisions} between particles.

In the frame of this kinetic description, and even though the
classical and academic ``two-level atom'' case is the simplest one may
consider, the general problem would consist in the self-consistent
solution of four coupled kinetic equations for, respectively, the
atoms in their ground state ($F_1$), the atoms in their excited state
($F_2$), electrons ($F_e$) -- and finally the photons ($I$).

However, and in order to be able to make first steps into the
\emph{extremely} complex and cumbersome full non--LTE radiation
transfer problem, we shall proceed with several simplifying
assumptions. First, we shall consider that \emph{only} the velocity
distribution function (hereafter VDF) of \emph{excited} atoms i.e.
$F_2$, may depart from a Maxwell-Boltzmann distribution. As previously
discussed by Oxenius (1979), we shall consider the velocity distribution
of the ground-state atoms, and of the free electrons to remain
Maxwellian with the same temperature $T$. The latter is also supposed
to be low enough i.e., $kT \ll h \nu_0$, where $\nu_0$ is the central
frequency of the transition between the two atomic levels of our
model, so that: (i) the number density of excited atoms if much
smaller than the one of ground-state atoms i.e., $n_2 \ll n_1$ and (ii)
stimulated emission may be further neglected.

Then we shall consider a time independent problem, as for photons, as
well as a \emph{stationnary} and \emph{force-free} problem for the
atoms.  And finally, following the debate between Oxenius (1965, 1979)
and Huben\'y (1981), we shall also neglect the potential
\emph{streaming} (i.e., ${v} . \partial_{r} \equiv 0$) of these
excited atoms. In that frame, the Boltzmann equation for atoms resumes
to:

\begin{equation}
  C[F_2]=0 \, ,
\end{equation}
where $C$ is the collision operator.
  
\section{The kinetic equation of excited atoms}

Each of the three contributions to the collision operator can be
explicited now, following Oxenius (1986).

The first one involves the radiation field, and can be expressed as:

\begin{equation}
  \left( {{\delta F_2} \over {\delta t}} \right)_{\rm rad.} = n_1
  B_{12} J_{12}(\vec{v}) f_1 (\vec{v}) - n_2 A_{21}
  f_2(\vec{v}) \, ,
\end{equation}
where stimulated emission has been neglected, and $A_{21}$ and
$B_{12}$ are the usual Einstein coefficients for, respectively,
spontaneous emission and radiative absorption. We also introduced here
the \emph{normalized to unity} $f_i$ distribution, such as $F_i(\vec{v})=n_i
f_i(\vec{v})$.

The second one involves inelastic collisions and writes:

\begin{equation}
  \left( {{\delta F_2} \over {\delta t}} \right)_{\rm inel.} = n_1 n_e
  C_{12} f_1(\vec{v}) - n_2 n_e C_{21} f_2(\vec{v}) \, ,
\end{equation}
where the $C_{ij}$ are collisional (de)excitation coefficients, and
$n_e$ is the electron density. Note that one may also consider that
the electrons are \emph{not} necessarily characterized by a
Maxwell--Boltzmann velocity distribution.

Finally, and according to Oxenius (1986; see also Bhatnagar et
al. 1954), a good approximation for the elastic collision term is:

\begin{equation}
  \left( {{\delta F_2} \over {\delta t}} \right)_{\rm el.} = n_2 Q_2
  \left[ f^{\rm M}(\vec{v}) - f_2(\vec{v}) \right] \, ,
\end{equation}
where $Q_2$ is a velocity-changing elastic collision rate. After
Borsenberger et al. (1986), one may consider $Q_2 = n_1 v_{\rm th.}
\sigma_{\rm el.}$, where $v_{\rm th.}$ is the ``most probable''
velocity of the atoms, and $\sigma_{\rm el.}$ is an ``average cross
section'' for this class of elastic collision.

The scattering integral $J_{12}$ which appears in Eq.\,(5) is defined
as:

\begin{equation}
  J_{12}(\vec{v}) = \oint{ { {d\Omega} \over {4 \pi} } } \int{
    \alpha_{12}(\xi) I_{\nu}(z, \vec{\Omega}) d\nu} \, ,
\end{equation}
where $\alpha_{12}$ is the \emph{atomic} absorption profile, and
$I_{\nu}(z, \vec{\Omega})$ the usual specific intensity. At this stage
we ommit on purpose the potentiel dependence with depth $z$ of
$J_{12}$, since we shall stay away still, in this study, from the
frame of radiation transfer. The specific intensity (see our next
\S4.) is also dependent on the direction $\vec{\Omega}$ of the
radiation. All integrals $d\Omega$ mean an angular integration over
all possible directions of propagation of light, as is usual in the
field of radiation transfer. It is more important now to keep in mind
that this scattering integral is a function of the velocity of the
massive particles. Indeed, we have to take into account the
Fizeau--Doppler relationship between frequency $\xi$ in the atomic
frame, and $\nu$ in the observer's frame:

\begin{equation}
  \nu = \xi + \left( \frac{\nu_0}{c} \right) \vec{v} \cdot \vec{\Omega}
\end{equation}
where $\vec{v}$ is the velocity of the atoms, and
  $\vec{\Omega}$ is the direction of propagation of the photons. This
latter coupling, and the explicit dependence of $J_{12}$ on the
velocity of these ``scattering centers'' which constitute the atoms
present in the atmospheric plasma, lead also naturally to the role
that will be played by their velocity distribution.

The \emph{atomic} absorption profile $\alpha_{12}$ is known a
priori, and we may consider either coherent scattering, i.e.:

\begin{equation}
\alpha_{12}(\xi)=\delta(\xi - \nu_0) \, ,
\end{equation}
where $\xi$ is the incoming photon frequency, or radiation damping
which is characterized by a Lorentzian profile such that:

\begin{equation}
  \alpha_{12}(\xi) = \left( a \over \pi \right)
        { 1 \over {(\xi - \nu_0)^2 + a^2} } \, ,
\end{equation}
and where, in both cases, $\nu_0$ is the central wavelength of the
transition we consider. In the remainder of this study, we shall
assume coherent scattering in the atomic frame, together with
  ``infinitely sharp'' energy levels for the model spectral line at
  $\nu_0$.

It is finally important to keep in mind, at this stage, that the atomic
emission profile $\eta_{21}$ may \emph{differ} from $\alpha_{12}$, in
general. This, combined with possible deviations from a
Maxwell--Boltzmann VDF for excited atoms will ``naturally'' lead to
\emph{non standard} as well, and \emph{radiation dependent}
redistribution in frequency for the photons.
  
\section{The kinetic equation of the photons}

The radiation transfer equation (RTE) is also the kinetic equation
governing the distribution, or specific intensity $I$, of the
photons. The correspondance between these two ``forms'' is nicely
described in \S3. of Oxenius (1986).

Assuming the very usual 1D plane--parallel geometry, the \emph{time
  independent} RTE can be written\footnote{As a matter of fact, for
  the respective kinetic equations of atoms and photons to be
  equivalent, while being further solved self--consistently, the issue
  of the streaming of particles described by a spatial derivative
  term, which remains further in the RTE but in the kinetic equation
  for the excited atoms, has to be discussed carefully (see again
  Huben\'y 1981, and references therein).} as:

\begin{equation}
    \frac{\partial I_{\nu}(z, \vec{\Omega})}{dz} = \kappa_{\nu} (z,\vec{\Omega})
    \left[ S_{\nu}(z, \vec{\Omega}) - I_{\nu}(z, \vec{\Omega}) \right]
 \end{equation}
where $\kappa_{\nu} (z,\vec{\Omega})$ is the absorption coefficient,
and $S_{\nu}(z, \vec{\Omega})$ the so-called source function, which
  a priori depends on depth $z$ in the atmosphere, frequency $\nu$ and
  direction $\vec{\Omega}$.

The absorption coefficient is usually defined as:

\begin{equation}
  \kappa_{\nu}(z, \vec{\Omega}) = \left( { {h \nu_0} \over {4 \pi} } \right)
  \left[ n_1 B_{12} \varphi_{\nu}(\vec{\Omega}) - n_2 B_{21}
    \psi_{\nu}(\vec{\Omega}) \right] \, ,
 \end{equation}
where $B_{21}$ is the Einstein coefficient for stimulated emission,
such as $g_1 B_{12}=g_2 B_{21}$ assuming the statistical weights
$g_i$, and the (line) source function is, by definition, the ratio
between the emissivity:

\begin{equation}
  \eta_{\nu}(z, \vec{\Omega}) = \left( { {h \nu_0} \over {4 \pi} }\right) n_2
  A_{21} \psi_{\nu}(\vec{\Omega}) \, ,
 \end{equation}
and the absorption coefficient. And finally, the more usual RTE
introduces the optical depth $\tau_{\nu}(\vec{\Omega}) =
-\kappa_{\nu}(z, \vec{\Omega}) dz$ in order to substitute the
geometrical length for another quantity more relevant to what photons
effectively experience during multiple light--matter interactions.

Should we neglect \emph{completely} stimulated emission i.e., going
beyond the (very usually) adopted so-called ``weak radiation field
regime'' for which stimulated emission is treated as ``negative''
absorption (as in Eq.\,13), then the source function can be simply
written as:

\begin{equation}
  S_{\nu}(z, \vec{\Omega}) = { {n_2 A_{21} \psi_{\nu}(\vec{\Omega})} \over {n_1
      B_{12} \varphi_{\nu}(\vec{\Omega})} } \, .
 \end{equation}
This assumption is also consistent with the low--temperature
assumption, which is also compatible with $n_1 \gg n_2$.

Furthermore, the respective (and potentially also depth-dependent)
emission and absorption profiles are defined as:

\begin{equation}
    \varphi_{\nu}(\vec{\Omega}) = \int{
      \alpha_{12}(\xi) f_1(\vec{v}) d^3 \vec{v}
    } \, ,
 \end{equation}
and,

\begin{equation}
    \psi_{\nu}(\vec{\Omega}) = \int{
      \eta_{21}(\xi, \vec{\Omega}) f_2(\vec{v}) d^3 \vec{v}
    } \, .
 \end{equation}
In the latter expressions appear explicitly the respective VDFs for
these atoms on the ground state, $f_1$, and for the ones in their
first excited state, $f_2$.

In general, atomic profiles $\eta_{21}$ and $\alpha_{12}$ are
\emph{not} identical. Moreover, the emission profile $\eta_{21}$ will
depend on the radiation field. We shall return to this issue in our
section 8.

\section{A two--distribution, two-level atom problem}

Hereafter we shall therefore simplify the problem, considering that
\emph{only} $f_2$ and the specific intensity $I$ may depart from
equilibrium distributions.

We shall first reassess earlier results which may still be quite
largely unknown to the astronomical community though, by considering
coherent scattering i.e., $\alpha_{12}(\xi)=\delta(\xi-\nu_0)$, where
$\nu_0$ is the central wavelength of the spectral line associated with
our two-level atom.

Let us therefore start with the additional simplifications of, a
stationnary, force-free and no advection (or streaming) case for
which:

\begin{equation}
  \left( {{\delta F_2} \over {\delta t}} \right)_{\rm rad.}
  + \left( {{\delta F_2} \over {\delta t}} \right)_{\rm inel.}
  + \left( {{\delta F_2} \over {\delta t}} \right)_{\rm el.} = 0 \, .
\end{equation}
    
Under these assumptions, we can derive the following expression for
$f_2$ and, while adopting a Maxwellian velocity distribution for
electrons, using the more classic formalism such that: $C_{ij}
\leftarrow n_e C_{ij}$ -- see also Eq.\,(9.51) of Hubeny \&
  Mihalas (2014) for instance, we have:


\begin{widetext}
\begin{equation}
  n_2 (A_{21} + C_{21}) f_2(\vec{v}) + n_2 Q_2 [f_2(\vec{v}) -
    f^M(\vec{v})] =  n_1 (B_{12} J_{12}(\vec{v}) + C_{12})
  f_1(\vec{v}) \, .
\end{equation}
\end{widetext}

Should we assume that $f_{1,2}$ are both Maxwellian, which is a common
assumption even for non--LTE radiation transfer problems, and should
we integrate Eq.\,(19) over all velocities, we recover the classic
form of the equation of statistical equilibrium (ESE) for a two-level
atom case, neglecting stimulated emission:

\begin{equation}
  n_2 (A_{21} + C_{21}) =
  n_1 (B_{12} \int_{\vec{v}}{ J_{12}(\vec{v})
    f^{M}(\vec{v}) d^3 \vec{v}} + C_{12})  \, .
\end{equation}
We shall come back soon to the integral term after Eq.\,(30).  Note
here that, because this particular contribution vanishes after
integration over velocities, the potential effects of
velocity-changing elastic collisions can be considered explicitly and
reasonably discussed though, \emph{only} by using such a kinetic
description.

Now from Eq.\,(19), we are able to derive an \emph{explicit}
expression for the \emph{non}--equilibrium VDF $f_2$, assuming that
all atoms in their fundamental state of energy follow a Maxwellian VDF
i.e., $f_1 \equiv f^{M}$. As in Borsenberger et al. (1986), we shall
normalize the specific intensity to the ``low temperature limit'' of
the Planck function, that is the Wien function:

\begin{equation}
  {\cal{B}}_W = \left( \frac{2h\nu_0^3} {c^2} \right) e^{-h\nu_0/kT}  \, .
\end{equation}
The use of this expression, instead of the Planck function is also
consistent with our assumption of a negligible rate of stimulated
emission. We shall also define the ratio:

\begin{equation}
  \bar{J}_{12} = {J}_{12}/{\cal{B}}_W  \, ,
\end{equation}
as well as use the following relationships, first between Einstein
coefficients:

\begin{equation}
  \frac{A_{21}}{B_{21}} = \left( \frac{2h\nu_0^3} {c^2} \right)  \, ,
\end{equation}
with $g_1 B_{12}=g_2 B_{21}$, and where $g_{1,2}$ are the statistical
weights of each level.

In addition, collisional rates are such that:

\begin{equation}
  n_1^* C_{12} = n_2^* C_{21} \, ,
\end{equation}
where $n_{1,2}^*$ are the LTE values of the respective densities of population,
satisfying the Boltzmann law:

\begin{equation}
  \left( \frac{n_2^*}{n_1^*} \right) = \left( \frac{g_2}{g_1} \right)
  e^{-h\nu_0/kT} \, .
\end{equation}
We shall therefore now introduce the \emph{normalized} population, which will
allow some measurement of deviations from LTE: $\bar{n}_2=n_2/{n_2^*}$.

Let us now introduce:

\begin{equation}
  \varepsilon = \frac{C_{21}} {A_{21} +  C_{21}}  \, ,
\end{equation}
which is another parameter otherwise traducing the amount of ``departure from
LTE'' in a given atmosphere, and the less common:

\begin{equation}
  \zeta = \frac{Q_{2}} {A_{21} +  C_{21}}  \, ,
\end{equation}
which characterizes the amount of elastic collisions. Assuming that
${n_1}={n_1^*}$ and $f_1 \equiv f^{M}$, we may now rewrite the following:

\begin{widetext}
\begin{equation}
  \bar{n}_2 f_2(\vec{v}) - \bar{n}_2 \zeta[f^M(\vec{v}) -
    f_2(\vec{v})] = \left[ \varepsilon + (1-\varepsilon) \bar{J}_{12}(\vec{v})
    \right] f^{M}(\vec{v}) \, .
\end{equation}
\end{widetext}
This expression is indeed identical to Eq.\,(2.44) of Borsenberger et
al. (1986) when one neglects streaming of particles (his $\eta=0$).

Now, integrating Eq.\,(28) \emph{over all velocities}, we find that:

\begin{equation}
  \bar{n}_2 =  \varepsilon + (1-\varepsilon) {\cal{J}}_{12}  \, , 
\end{equation}
where we defined:

\begin{equation}
  {\cal{J}}_{12} = \int_{\vec{v}}{ \bar{J}_{12}(\vec{v})
    f^{M}(\vec{v}) d^3 \vec{v}} \, ,
\end{equation}
which is equivalent to the common $\bar{J}$ scattering integral used
in the standard approach. We finally recover the following:

\begin{widetext}
\begin{equation}
  f_2(\vec{v}) = \left[ \frac{\zeta}{1+\zeta} +
    \left(\frac{1}{1+\zeta}\right) \frac{\varepsilon + (1-\varepsilon)
      \bar{J}_{12} (\vec{v})} {\varepsilon + (1-\varepsilon)
      {\cal{J}}_{12}} \right] f^{M}(\vec{v}) \, .
\end{equation}
\end{widetext}
This is the central relationship allowing us to quantify departures
from ``Maxwellianity'' for $f_2$, and potential differences between
macroscopic emission and absorption profiles, depending on the nature
of the radiation field characterized by $I$.

\section{Two comments on the standard theory}

Complete redistribution in frequency relies on the a priori
  assumption that the ``macroscopic'' emission and absorption profile
  are identical. Then the line source function reduces to the simple
  ratio: $S=n_2 A_{21}/n_1 B_{12}$ i.e., a quantity \emph{independent}
  of the frequency after Eq.\,(15). It is also a \emph{considerable}
  simplification of the NLTE radiation transfer problem. Assuming, as
  we do here, that the atomic profiles are identical i.e.,
  $\alpha_{12} \equiv \eta_{21}$, CRD also assumes implicitly the
  identity:

\begin{equation}
f_2 \equiv f_1 ( \equiv f^{M} )  \, ,
\end{equation}
whose effects therefore result from expressions given in
  Eqs.\,(16) and (17). In that frame, we also have that the
normalized source function $\bar{S}=S/B_W$ is such that:

\begin{equation}
  \bar{S} = \bar{n}_2 = \varepsilon + (1-\varepsilon) {\cal{J}}_{12} \, .
\end{equation}
This expression is indeed consistent with the classical (CRD)
``two-level atom'' case described in any radiation transfer textbook,
since Avrett \& Hummer (1965).

However, eliminating the case of little astrophysical interest of
  an isotropic ``white'' radiation field, Eq.\,(32) is clearly
\emph{inconsistent} with the more general Eq.\,(31), unless setting
$\zeta=0$ \emph{and} $\varepsilon=1$ i.e. assuming LTE! Therefore
non--LTE should \emph{always} come together with some partial
frequency redistribution, and no complete redistribution a priori.

Additionally, even when ``standard'' partial frequency redistribution is
used, with \emph{a priori known} redistribution functions, in most cases a
``branching ratio'' usually defined as:

\begin{equation}
  \gamma = \frac{A_{21} + C_{21}}{A_{21} + C_{21} + Q_E}
\end{equation}
is introduced (see Omont et al. 1972, Heinzel \& Huben\'y 1982,
Heinzel et al. 1987), to form a linear combination of $R_{\rm II}$ and
$R_{\rm III}$ redistribution such as:

\begin{equation}
  R(x',x) = \gamma R_{\rm II}(x',x) + (1-\gamma) R_{\rm III}(x',x) \ ,
\end{equation}
following the nomenclature of Hummer (1962), and where $x'$ and $x$
are, respectively, the usual incoming and outgoing reduced
frequencies, usually defined as:

\begin{equation}
x=\frac{\nu - \nu_0}{\Delta \nu_D} \, ,
\end{equation}
with the Doppler width $\Delta \nu_D = (\nu_0/c) {v}_{th.}$, and the most
probable velocity ${v}_{th.}=\sqrt{2kT/M}$ as usual.

The elastic collision rate $Q_E$ which appears in the branching
  ratio $\gamma$ should \emph{not} be confused with the $Q_2$ elastic
  collision rate introduced in Eq.\,(7). The latter is indeed, and
  more precisely the rate of these \emph{velocity-changing} elastic
  collisions i.e., a fraction of the total amount $Q_E$ of elastic
  collisions. Bommier (2016a), for instance, provides a synthetic but
  accurate and valuable explanation about the different classes of
  elastic collisions. Moreover, as mentioned in this discussion, Landi
  Degl'Innocenti \& Landolfi (2004; see their \S13.2) estimate that $Q_2$
  should be very small in solar--like atmospheres, so that the
  so-called Boltzmann elastic collision term may be practically
  neglected for radiative modeling under such physical conditions.

  Nevertheless, quantifying any significant effet due to elastic
  collisions in the NLTE problem for other atmospheres with physical
  conditions characterized by a more important amount of
  velocity-changing collisions than in the solar--like case, would
  thus require using the more general theoretical frame we adopt here
  (see also Belluzzi et al. 2013 for instance). Indeed, using velocity
  integrated expressions such as Eqs.\,(20) or (29), which is the case
  in the standard NLTE approach, and even with standard PRD though,
  the elastic collision term introduced in Eq.\,(7) just
  \emph{vanishes}.

\section{Quantifying departures from equilibrium}

The stage is now set for a first evaluation of potential departures
from ``Maxwellianity'', owing to the nature of the radiation field we
may adopt.

For a first exploration, we shall first assume pure Doppler broadening
for the spectral line, that is to say coherent scattering in the
atom's frame and:

\begin{equation}
  \alpha_{12}(\xi) = \eta_{21}(\xi) = \delta(\xi-\nu_0)  \, ,
\end{equation}
where both atomic profiles are also assumed to be \emph{isotropic}.
Should we moreover assume {\emph no} elastic collisions so far i.e.
$\zeta=0$, then after Eq.\,(31) we end-up with:

\begin{equation}
  f_2(\vec{u}) = \left[ \frac{\varepsilon + (1-\varepsilon)
      \bar{J}_{12} (\vec{u})} {\varepsilon + (1-\varepsilon)
      {\cal{J}}_{12}} \right] f^{M}(\vec{u}) \, ,
\end{equation}
where we also introduced the \emph{normalized} velocity
$u=v/v_{th.}$.

\begin{figure}[]
  \includegraphics[width=9.5 cm, angle=0]{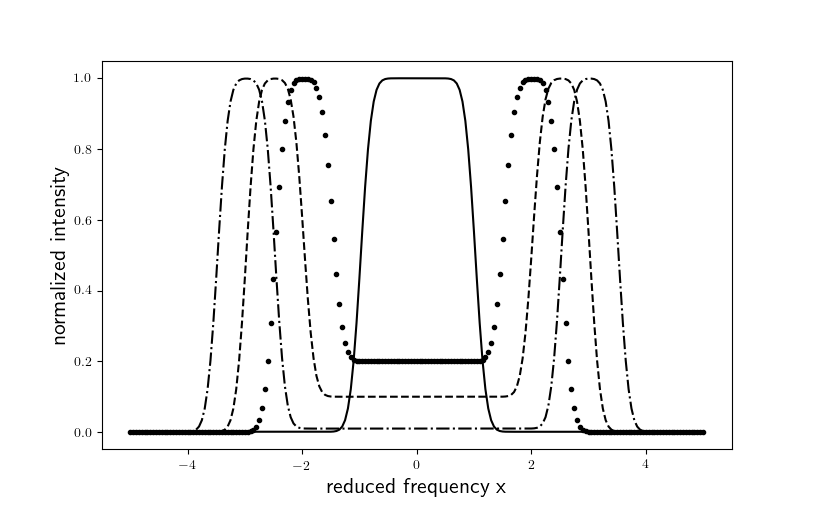}
  \caption{Isotropic and normalized intensity profiles $I$ vs. the
    reduced frequency $x$, used for the evaluation of departures from
    the equilibrium velocity distribution for the VDF of these excited
    atoms, $f_2$. We adopted different values for (i) the location of
    the (symmetric) peaks, and (ii) of the central reversal of each
    profile. These variations can also mimic some resonance lines
    observations that may form under different astrophysical
    conditions.}
  \label{Fig1}
\end{figure}

Non--LTE implies that $\varepsilon<1$, which in turn
\emph{necessarily} implies that $f_2$ differs from being Maxwellian in
general. Should these differences be significant enough, then after
Eq.\,(17) and despite the equality between respective atomic profiles,
we however expect differences between macroscopic absorption and
emission profiles, unlike what assumes \emph{ab initio} the widely
adopted CRD.

The abovementioned assumptions lead to:

\begin{equation}
      \bar{J}_{12} (u) = \frac{1}{2u} \int_{-u}^{+u} {I(x) dx}  \, .
\end{equation}
Then, since $f_1 \equiv f^M$, using this classical expression for the
(normalized to unity) macroscopic absorption profile:

\begin{equation}
      \varphi_x = \frac{1}{\sqrt{\pi}} e^{-x^2}  \, ,
\end{equation}
we can also establish that:

\begin{equation}
      {\cal{J}}_{12} = \frac{1}{\sqrt{\pi}} \int_{-\infty}^{+\infty}
      {I(x) e^{-x^2} dx} \, ,
\end{equation}
which finally allows to evaluate departures of $f_2$ from the
equilibrium distribution depending on the thius far given (and
normalized to unity) radiation field $I(x)$.

\begin{figure}[]
  \includegraphics[width=9.5 cm,angle=0]{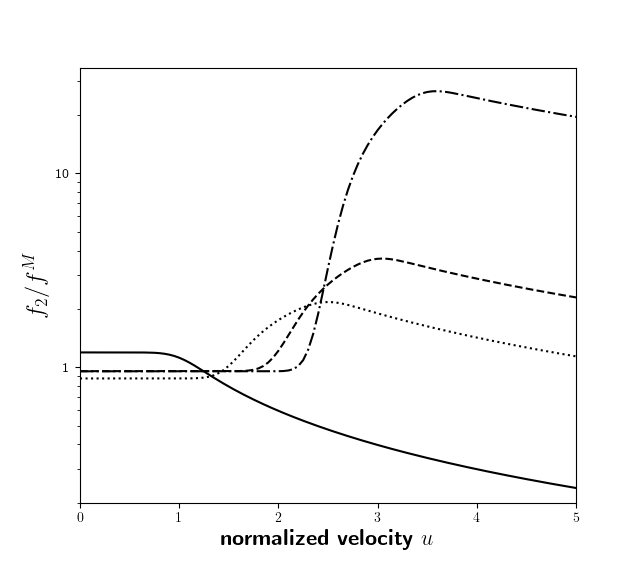}
  \caption{Ratios $f_2/f^M$ as a function of the normalized velocity
    $u$, for different radiation field profiles, and for
    $\varepsilon=0$ and $\zeta=0$. The same linestyle convention as in
    Fig.\,(1) was adopted. The magnitude of these departures from a
    Maxwellian velocity distribution depends on both (i) the distance
    of the peaks of $I$ vs. line center, and (ii) the amount of its
    central reversal characterized by $I_{\rm max.}/I(0)$.}
  \label{Fig2}
\end{figure}

According to these expressions, one can notice that, with coherent
scattering in the atom's frame, it remains however a \emph{ultimate}
option for ``saving'' the CRD case. Indeed, assuming isotropic
``white'' light illumination i.e., a frequency independent radiation
field such as $I(x)=I_0$ is the only condition left to us in order to
keep $f_2 \equiv f^M$, and therefore $\psi_x \equiv \varphi_x$.

Further illustrative computations have been made using several
incident radiation profiles, as displayed in Fig.\,(1). Our selection
covers distinct possibilities of central reversal amplitudes i.e.,
$I_{\rm max.}/I(0)$, as well as peaks located at different, and
increasing positions away from line center. These may mimic some
observed and well-resolved resonance line profiles, similar to one of
the good candidates for future modeling such as Ly$\alpha$ of H\,{\sc
  i}.

Then, we computed ratios of $f_2/f^M$ according to Eq.\,(38), and
adopting further that $\varepsilon=0$. Our results for $f_2/f^M$ as a
function of the normalized velocity $u$ are displayed in Fig.\,(2), for
which we used the same linestyle convention as in Fig.\,(1).

Departures from a Maxwellian distribution show up at almost all
normalized velocities $u$. These departures tend to increase both with
(i) an increasing distance of the peaks of the incident radiation
profile away from line center, as well as with (ii) an increase of the
central reversal of the incident profile, which is characterized by
the ratio between the maximum value of the profile $I_{\rm max}$ and
its central value $I(0)$.

Finally, it is also possible to compute the ratio between the
macroscopic emission profile $\psi_x$ and the (Gaussian) absorption
profile $\varphi_x$. Indeed, after Eqs. (38) and (17), and with
$\varepsilon=0$ though, one can derive (see also Appendix A) that:

\begin{equation}
  \psi_x = \frac{2}{\sqrt{\pi} {\cal{J}}_{12}} \int_{\lvert x \rvert}^{\infty}
      {\bar{J}_{12}(u) u e^{-u^2} du}  \, ,
\end{equation}
for the case of coherent scattering in the atom's frame.

\begin{figure}[]
  \includegraphics[width=9.5 cm,angle=0]{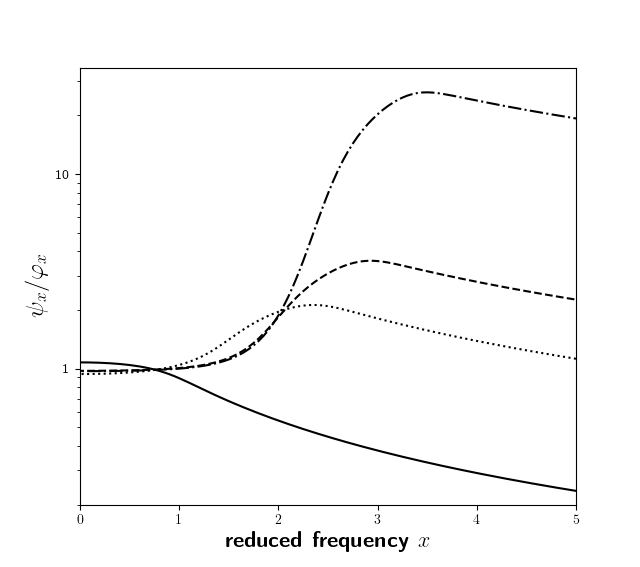}
  \caption{Ratios $\psi_x/\varphi_x$ as a function of the reduced
    frequency $x$, for the different radiation field profiles
    displayed in Fig.\,(1), using also the same linestyle
    convention. We observe here, for the case of coherent scattering
    in the atom's frame, similar tendencies as the ones displayed in
    Fig.\,(2) for the ratios between VDFs.}
  \label{Fig3}
\end{figure}

Note also that, at this stage, with the knowledge of the
\emph{radiation dependent} emission profile, we are also able to
compute the source function, following Eq.\,(15).

Significant differences between emission and absorption profiles are
put in evidence from our relatively simple computations. They are
displayed in Fig.\,(3) which shows macrocopic profiles ratios as a
function of $x$, for the same set of incident radiation profiles as
before.

We observe for these ratios, and the special case of coherent
scattering in the atom's frame, the same tendancies as the ones
already observed for the ratio between VDFs. It essentially shows the
very limit of the theoretical framework of CRD, and its intrinsic
\emph{inconsistency} with NLTE.

\begin{figure}[]
  \includegraphics[width=9.5 cm,angle=0]{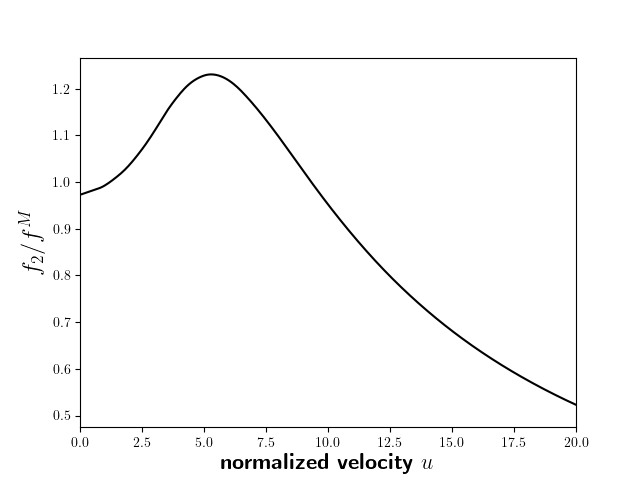}
  \caption{Ratio $f_2/f^M$ vs. the normalized velocity $u$
    for $\varepsilon=10^{-7}$ ($\zeta=0$), and a realistic H\,{\sc i}
    Ly$\alpha$ illumination profile as observed by SoHO/Sumer. The peak of
    the line profile is located $x \sim 5$, and the central
    self--reversal is about 0.7 for this resonance line (see Fig.\,5
    of Gun\'ar et al. 2020).}
  \label{Fig4}
\end{figure}

\section{Discussion}

So far, the radiation field has been imposed a priori using a simple
$I(x)$ function. Potential anisotropies of the radiation field could
be taken into account within this formalism, of course. Also, more
realistic line profiles will be adopted further, as we did for the
computation of the ratio $f_2/f^M$ displayed in Fig.\,(4). In that
case, we used the profile of the H\,{\sc i} Ly$\alpha$ resonance line
deduced from SoHO/Sumer solar observations (Gun\'ar et al. 2020, in
particular their Fig.\,5). We adopted a temperature of $10^4$ K
leading to a $I(x)$ profile extending up to $x\sim30$. Then we fixed
$\zeta=0$ and $\varepsilon=10^{-7}$ i.e., a value which is quite
typical of conditions usually met in solar prominences for instance
(see e.g., Paletou et al. 1993). Departures from a Maxwellian
distribution mainly appear significant around the peak of the line
profile at $x \sim 5$ in our case, and in the line wings.

But most important will be to address further the self--consistent
computation of the two distributions $f_2$ and $I$. This should
be possible within an iterative cycle involving: an initial guess for
the source function $S$ assuming for instance both LTE and CRD, then a
\emph{formal solution} of the RTE allowing for the computation of
$\bar{J}_{12}$ and ${\cal{J}}_{12}$. Then $f_2$ will be estimated, as
well as $\psi_x$, making it possible to update further the source
function, start another iteration and so forth, until reaching
convergence for \emph{both} $f_2$ and $I$.

We believe that state-of-the-art iterative methods presently used for
NLTE radiative transfer (see e.g., Lambert et al. 2016, and references
therein) will allow one to explore all of the potential of this
alternative and more detailed theoretical frame, beyond what could be
achieved at that time by Borsenberger et al. (1986, 1987) and
Atanackovi$\check{\rm c}$ et al. (1987).

In parallel with this numerical set-up, we shall also consider the
more realistic case of ``natural broadening'' of the upper level of
the atomic transition. The alternative atomic absorption profile is
now a Lorentzian profile, whose expression was given in
Eq.\,(11). However, in such a case, atomic absorption and emission
profiles will be \emph{different} since the atomic emission profile
will now depend explicitely on the radiation field. The expression for
this emission profile $\eta_{21}$ is discussed in Appendix B of
Oxenius (1986); see also Huben\'y et al. (1983; their \S4.3). The
derivation of the latter expressions remains however questionable, and
it more likely leads to \emph{erroneous} coefficients (V. Bommier,
private communication). Therefore, the modified emissivity consisting
in the sum of a ``classical'' (order-2) term with an extra ``order-4''
contribution proposed by Bommier (1997) will be preferred. In that
frame, based on sound physical grounds, it appears that \emph{only}
the order-2 term emissivity would be associated with the iterative
modifications of the VDF of the excited atoms. Indeed, the so-called
order-4 coefficient introduced by Bommier (1997) involves the
\emph{lower level} atomic state -- see her Eq.\,(92) for the
unpolarized case. However, as well as the order-2 contribution (via
the statistical equilibrium equation), the order-4
contribution\footnote{This very contribution is also reminiscent of
  $(R_{\rm II} - R_{\rm III})$, which clearly appears looking at
  Eq.\,(102) of Bommier (1997), for the special case of an infinitely
  sharp lower atomic level.} to emissivity depends on the radiation
field, so that it will \emph{also} need to be ``runned to
convergence'' together with the distributions $I$ and $f_2$. The
associated numerical burden will mainly amount to recomputing
Voigt-like functions (see e.g., Paletou et al. 2020) at every
iteration, according to changes in $f_2$ and $I$, and at every depth
in the atmosphere.

Finally, any significant effect of these velocity-changing
\emph{elastic} collisions, which are ``by construction'' though not
taken into account in the standard non--LTE approach could be properly
discussed for different astrophysical situations, using the present
formalism.

\section{Conclusion}

We have revisited and rewritten, using more conventional notations
basic elements of Oxenius' kinetic approach for non--LTE spectral line
formation. We also brought clarification on some technical parts which
were not always clearly explained in his textbook and some other
subsequent, but still remaining ``little aknowledged'' studies.  Note
that similar elements of this approach have been recently discussed
again, in the more complete but more likely less accessible frame of
\emph{polarized} radiation transfer, using the density matrix
formalism (see e.g., Belluzzi et al. 2013).

Here we fully agree with the statement about \emph{``the necessity of
  resolving the statistical equilibrium equations for each velocity
  class''} expressed by Bommier (2016b; see her \S5.3).  Therefore, in
a next step, we shall implement an iterative numerical scheme, based
on pre-existing practices in astrophysical radiation transfer (see
e.g., Lambert et al. 2016 and references therein) in order to compute
\emph{self--consistently} both for the specific intensity (or
distribution of the photons) and, in a first stage, for the velocity
distribution of these excited atoms present in the atmosphere. As a
matter of fact, directly tackling the computation of the VDF of
excited atoms also naturally leads to a generalization of the PRD
theoretical frame, somewhat \emph{without} using any a priori defined
redistribution function.

This will allow us to identify further and explore in more detail
plausible astrophysical situations for which this ``full--NLTE''
approach for the very general issue of spectral line formation will
prove to be indispensable.

\begin{appendix}

  \section{Derivation of the macroscopic emission profile $\psi_x$}

    We find it quite unfortunate that in his monograph Oxenius
    (1986) did not make the derivation of the macroscopic
    emission profile whose expression is given in Eq.\,(42). This is
    why we propose the following steps.

  We shall first assume that:

  \begin{equation}
  \oint{ \delta(x - \vec{u} \cdot \vec{\Omega}) d\Omega = \frac{2 \pi}{u}
    H(u-|x|)}  \, ,
  \end{equation}
  where $H$ is the usual Heaviside, or unit step function. Using
    such an angular integration for the combination of our Eqs.\,(17)
    and (38) when $\varepsilon=0$ writes as:

  \begin{equation}
    \psi_x = \frac{1}{{\cal{J}}_{12}} \oint{ \left(
      \int{ \bar{J}_{12}(\vec{u}) \delta(x - \vec{u} \cdot \vec{\Omega})
        f^{M}(\vec{u}) d^3 \vec{u} } \right)
        \frac{d\Omega}{4\pi} } \, .
  \end{equation}
  %
%
First, the angular integration will transform the Dirac
  distribution modeling coherent scattering, into the Heaviside
  function.

Then, using spherical coordinates in the \emph{velocity
  space} for the integration in $u$, that is:

  \begin{equation}
    d^3 \vec{u} = u^2 \sin(\theta) du d\theta d\varphi \, ,
\end{equation}
  where $\theta$ and $\varphi$ are the usual, respectively, polar
  angle and azimuth, one gets:

    \begin{equation}
    \psi_x = \frac{2\pi}{\pi^{3/2} {\cal{J}}_{12}} \int{
      \bar{J}_{12}(u) u e^{-u^2} H(u-|x|) du} \, ,
    \end{equation}
    which finally leads to Eq.\,(42) using the property of the $H$
    function which is then different from 0 for $u \ge \lvert x
      \rvert $.

\end{appendix}

\begin{acknowledgements}
We are grateful to Dr Stanislav Gun\'ar (Astronomical Institute of the
Czech Academy of Sciences) who very kindly made his data available to
us, and to Dr V\'eronique Bommier (Cnrs and \emph{Observatoire de
  Paris}, France) for fruitful discussions and her
encouragement. Finally, Fr\'ed\'eric Paletou is grateful to his
radiative transfer {\it Sensei}, Dr. L.H. ``Larry'' Auer, with whom we
started discussing about these issues long time ago.
\end{acknowledgements}

\end{document}